\documentclass[prl,twocolumn,superscriptaddress]{revtex4}

\usepackage{amsmath}
\usepackage{bm}

\usepackage[dvips]{graphicx}
\usepackage{color}

\begin{document}

\title{Bubble Statistics and Dynamics in Double-Stranded DNA}
\author{B. S. Alexandrov} 
\affiliation{Theoretical Division, Los Alamos National Laboratory, Los Alamos, New Mexico 87544}
\affiliation{Physics Department, Florida Atlantic University, Boca Raton, Florida 33433}
\author{L.T. Wille}
\affiliation{Physics Department, Florida Atlantic University, Boca Raton, Florida 33433}
\author{K.\O. Rasmussen}
\author{A.R. Bishop}
\author{K.B. Blagoev}
\affiliation{Theoretical Division, Los Alamos National Laboratory, Los Alamos, New Mexico 87544}  
%\date{\today}
\pacs{63.20 Pw;87.15. He}

\begin{abstract}
The dynamical properties of double-stranded DNA are studied in the framework of the
Peyrard-Bishop-Dauxois model using Langevin dynamics. Our simulations are analyzed in terms of two 
probability functions describing coherently localized separations ('bubbles') 
of the double strand. We find that the resulting bubble distributions are more sharply peaked at the active sites than found in
 thermodynamically obtained distributions. Our analysis ascribes this to the fact that the {bubble} 
life-times significantly affects 
{the} distribution function. We find that certain base-pair sequences promote long-lived bubbles and we argue {that} this is 
due to a length scale competition 
between  the  nonlinearity and disorder present in the system. 

\end{abstract}

\maketitle

The role of dynamics in biological function is becoming increasingly clear \cite{Volkman_2001,Eisenmesser_2002,Tomschik_2005}.
Whereas protein action and binding have traditionally  been discussed in terms of 
static structures, it is now evident that many functionalities are consequences of 
dynamics. Because of its biological importance and structural clarity DNA constitutes an appropriate system in which to begin to understand 
how structure and thermal motion can work together to determine function.
In particular, the identification of biological processes that are regulated by the dynamical properties of 
DNA is fundamentally important for understanding its interaction with other molecules. 
Key mechanisms are controlled by entropically driven thermal fluctuations, which cause local dynamical 
changes in the inter-strand separation ('bubbles') in double-stranded DNA molecules. 
Recent theoretical and experimental studies \cite{Choi_2004} suggest that the 
base pair sequence (structure) determines specific regions in the double-strand that are 
more prone to such thermally induced strand separation. Most importantly, these studies 
have demonstrated a strong correlation between the specific location of large, coherent openings, in  DNA and transcription-promoting regions of the DNA 
sequence for several well-characterized viral sequences.  
It has also be found \cite{Blagoev} that the DNA dynamics 
in the presence of UV induced dimers between two adjacent thymine bases
(TT-dimers) is dramatically altered in the neighborhood of the dimer,
suggesting an enhancing role for the large fluctuations present at the dimer
site in the dimer recognition pathway. In both cases the theoretical characterization has been provided by the Peyrard-Bishop-Dauxois (PBD) 
model \cite{Peyrard_1989, Dauxois_1993}. However, recently it was argued \cite{vanErp,Zoi} that 
thermodynamic characterization of the thermal fluctuations may differ from a dynamical characterization, which points to the  
need for a thorough understanding of the dynamical effects in this highly nonlinear and cooperative material.

Here, we use finite temperature Langevin simulations to probe the impact of sequence heterogeneity on bubble dynamics 
in six different sequences all composed of 69 base pairs:
i) two homogeneous sequences composed purely of thymine-adenine (T-A) and guanine-cytosine (G-C) base-pairs, respectively, and 
 ii) two specific heterogeneous sequence: Adeno Associate Viral  (AAV) P5 promoter and a mutated (AAV) P5 promoter obtained from 
the wild  AAV promoter by replacing two specific neighboring A-T base-pairs with G-C base-pairs (see Ref. \cite{Choi_2004} for details). 
Finally, iii) we investigate two periodic sequences each containing 35 (T-A) base-pairs and 34 (G-C) base-pairs that have different periodicities - $G_{1}A_{1}$ - 
with a period of 2 base pairs and $G_{5}A_{5}$ with a period of 10 base pairs.  We compare our results to thermodynamic results for the inter-strand 
DNA opening obtained with the same model and observe several important differences. We emphasize that the Langevin dynamics of the PBD model's principal degrees-of-freedom
for DNA base pairs is necessarily a phenomenological representation of DNA's full complexity: microscopic fine scales of, e.g., water motions
are not explicitly modeled.

In the framework of the PBD model, the thermal dynamics of the $n$'th base-pair
is obtained through:
\begin{eqnarray}
\label{eqm2}
m \ddot y_n = &-& V^\prime (y_n) - W^\prime (y_{n+1},y_{n}) 
- W^\prime (y_{n},y_{n-1})\nonumber \\ &- &m\gamma \dot y_n +\xi_n(t), 
\end{eqnarray}
where 

$V(y) = D_{n} (e^{- a_{n} y}-1)^2,$

is an on-site Morse potential modeling the hydrogen bonding of complementary bases, and representing the 
exact sequence \cite{sequence}, and

$W(x,y)= \frac{k}{2} (1+\rho e^{-\beta(x+y)})(x-y)^2 $

represents the nonlinear stacking interactions. Here
prime denotes differentiation with respect to $y_n$, $\gamma$
is the friction constant, and the random force $\xi_n(t)$ is Gaussian distributed white noise.
With the  parameter values used (see Refs. \cite{Campa,gamma}, the success of the model in describing the base-pair openings of 
double-stranded DNA has been demonstrated by direct comparison with various 
experiments on the melting transition \cite{Campa}, S1 nuclease digestion \cite{Choi_2004}, pre-denaturation 
bubbles \cite{Ares_2005}, and forced unzipping \cite{Voulgarakis}.

\begin{figure}[h]
   \includegraphics[width=\columnwidth]{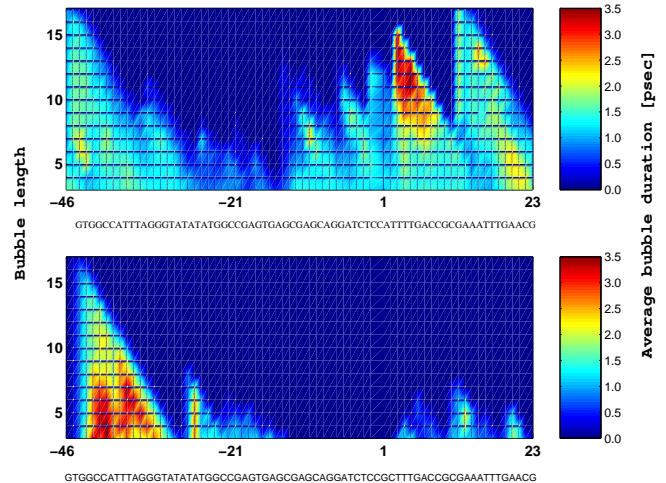}
   \caption{Average bubble duration time, $\sum_{tr=4.5 \AA}^ \infty ABD(n,l,tr)$, for  AAV P5 (upper) and for mutated AVV P5 (lower)  
sequences. \label{fig:4}}
\end{figure}

Here we simulate the dynamics of double-stranded DNA at $T=300$K by numerically integrating the system 
of stochastic equations (\ref{eqm2}), applying
periodic boundary conditions. In the presence of the thermal bath,  modeled by the random forces, the 
creation of a bubble is a stochastic process \cite{Liapunov}
most appropriately described in terms of a probability. 
We define the probability for bubble 
existence as:
\begin{eqnarray}
P_n(l,tr)=\left\langle \frac{1}{t_s}
\sum_{q_{n}^k=1}^{q_{n}^{k_{max}}(l,tr)}\Delta t[q_{n}^k(l,tr)] \right\rangle_M
\label{PDF1}
\end{eqnarray}
where $\left\langle{...} \right\rangle_M$ denotes averaging 
over $M (\sim 1000)$ simulations. The integer index, $q_{n}^k(l,tr)$, enumerates the bubbles defined as a double-strand separation of amplitude $tr\geq 0.5\AA$, spanning $l\geq 3$ consecutive base-pairs beginning at the $n$th base-pair in
the $k$th simulation. In practice we bin $P_n(l,tr)$ using bin sizes $l=1$  and $tr=0.5\AA$. The quantity $\Delta t[q_{n}^k(l,tr)]$ is the existence time of the $q_{n}^k(l,tr)$'th  bubble.  $t_s \sim 1~-~2\, $nsec is the  duration  of a single simulation (bubble life-times are in the picosecond range and are much smaller see Fig. \ref{fig:4} ). Probabilities 
for bubble existence at a given site, for all lengths and amplitudes defined in this way 
are obviously normalized  since all possible openings at every step of the simulation are counted.
\begin{figure}[h]
   \includegraphics[width=\columnwidth]{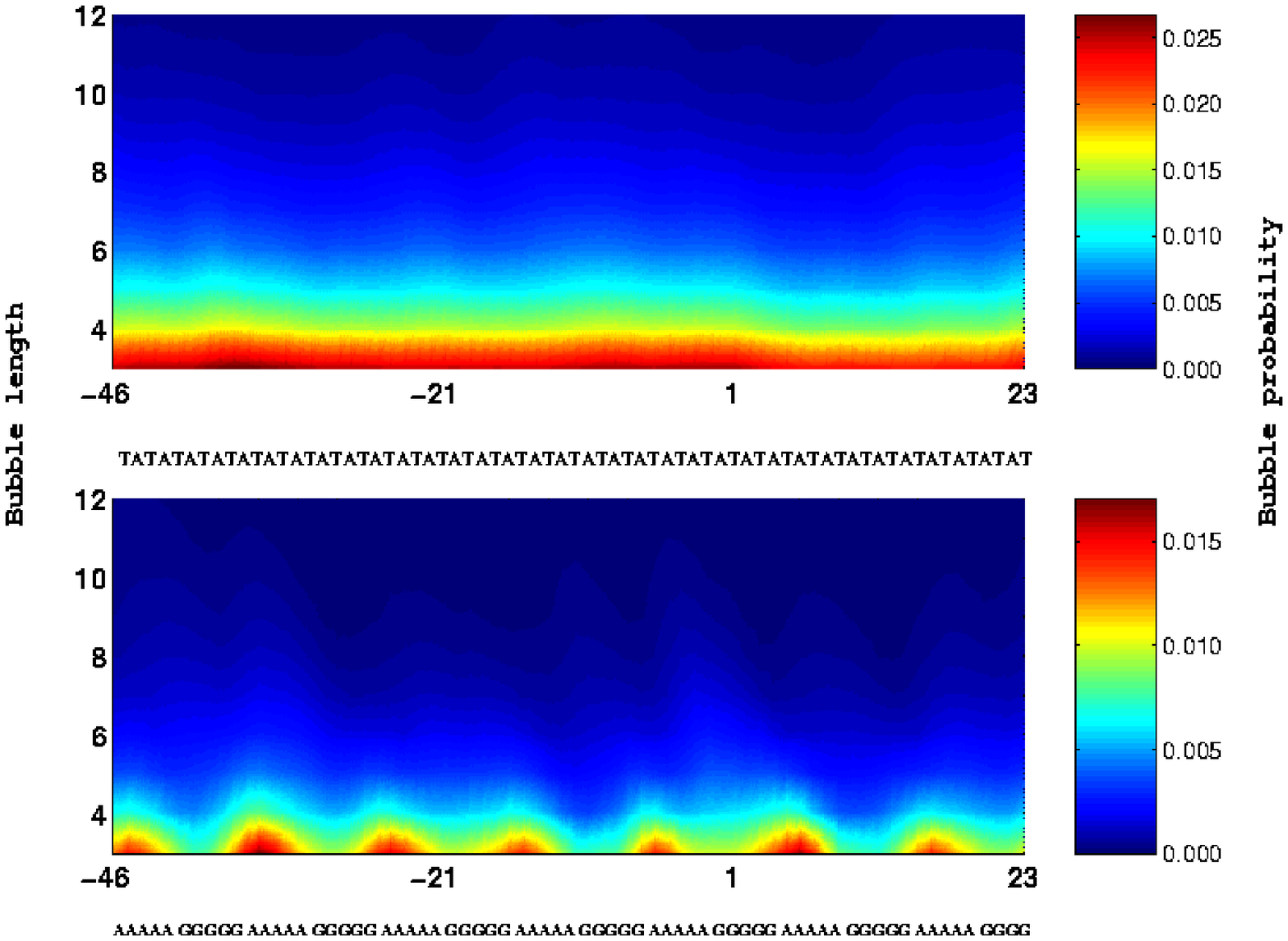}
   \caption{Bubble probability, $\sum_{tr=1.5 \AA}^\infty P_n(l,tr)$, for the 69 base-pair homogeneous $A-T$ sequence 
(upper) and for the periodic $G_5A_5$ sequence (lower).} \label{fig:1}
\end{figure}
The plot of $\sum_{tr=1.5 \AA}^\infty P_n(l,tr)$ as obtained from Eq. (\ref{PDF1}) given in Fig. \ref{fig:1}
demonstrates two clear results:

1) The probabilities for bubble formation in a homogeneous $T-A$ sequence ( or for a G-C sequence; not shown) do not depend on the base pair 
index because of the translation invariance. 

2) For the periodic sequence $G_{5}A_{5}$, the probabilities are periodic with the period of the sequence. We can 
clearly identify the {\it sources} of the bubbles to be situated in the AT-rich half-period of the sequence. In contrast, we observed
for a $G_{1}A_{1}$ sequence (not shown) the probabilities to be almost spatially uniform because of the short periodicity - only a single 
$G-C$ base pair between the two $A-T$ base pairs.  
The important observation from these results
is that not only the length of the hot spots (the $T-A$ areas)  but also the length of the ``barriers'' (the $G-C$ intervals) play crucial roles for 
the probability of the bubble existence by restricting, through impedance mismatch, the energy flow from the $A-T$ rich regions. Clearly bubbles of all 
length have thermodynamic weight, increasing with temperature and decreasing with bubble size. However, the base pair inhomogeneity preferentially selects long-lived
 bubbles of specific sizes at specific locations. This is a result of length scale competitions inherent in nonlinear, disordered systems \cite{Sanches_1988}. 
\begin{figure}[h]
   \includegraphics[width=\columnwidth]{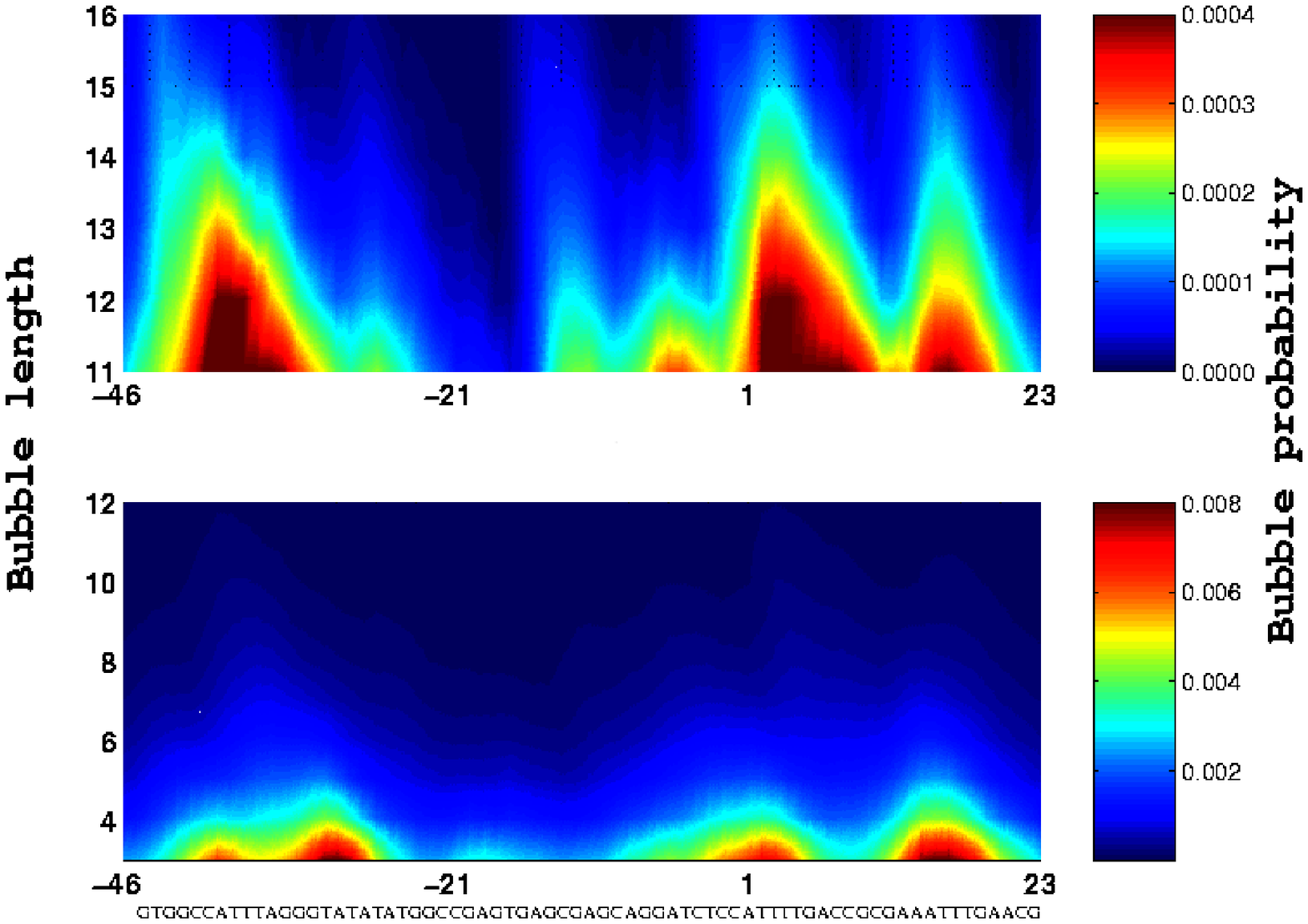}
   \caption{Bubble probability, $\sum_{tr=1.5 \AA}^\infty P_n(l,tr)$, for the 69 base-pair AAV P5 promoter (see Ref. \cite{Choi_2004}), with $t_s=2\, $nsec and 
$M=800$ simulations.}
\label{fig:2}
\end{figure}

In Fig. \ref{fig:2} we show $\sum_{tr=1.5 \AA}^\infty P_n(l,tr)$ 
for the 69 base-pair adeno associate viral (AAV) P5 promoter (see Ref. \cite{Choi_2004}). Two regions are prominent in terms of
 largest probabilities for bubbles existence.  These regions are located around base-pairs 
$+1$ and $-30$, which have previously been identified as the transcription start site (TSS), and the binding site for the TATA-binding 
protein (TBP), respectively. It is noteworthy that the probability becomes more localized around the 
identified sites as increasingly longer (in terms of consecutive sites) bubbles are considered. This result is in agreement 
with findings in similar simulations investigating the frequency of the base pair opening \cite{Choi_2004}. There is also 
good agreement with the thermodynamic results \cite{Zoi,vanErp} derived from the PBD model, however the active regions are much more 
sharply identified here than in the case of the thermodynamic treatment. 

{Our results confirm that the bubble localized at the transcription promoter site
can aid the RNA polymerase and the associated proteins in the formation of the transcription bubble}
 
It has been argued \cite{vanErp} that the results obtained in Ref. 
\cite{Choi_2004} are flawed by insufficient statistics in the simulations. We therefore present in Fig. \ref{fig:2} the result of 
$M=800$ simulations of $t_s= 2$nsec duration. Similar results were obtained from simulations of $t_s= 1$nsec duration (not shown) suggesting
that the statistics is indeed sufficient even in a 1 nsec simulation. It is important to note that even for the 2 nsec simulations 
we never observed complete melting of all the 69 base pairs indicating that we are exclusively sampling the premelting regime.
\begin{figure}[h]
   \includegraphics[width=\columnwidth]{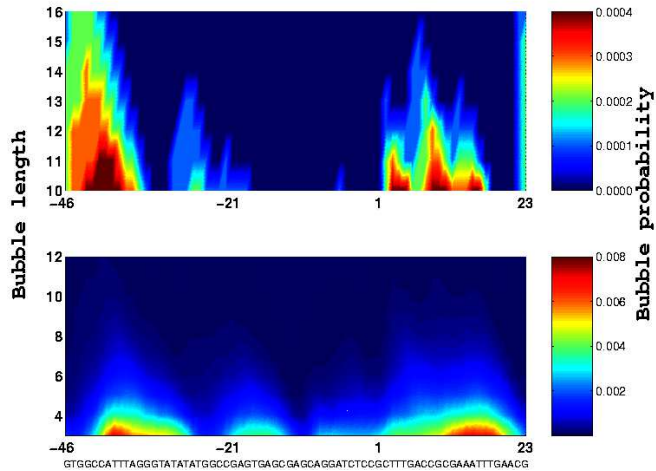}
   \caption{Bubble probability, $\sum_{tr=1.5 \AA}^\infty P_n(l,tr)$, for a mutated AVV P5 
sequence (see text). \label{fig:3}
   }
\end{figure} 

In Fig. \ref{fig:3} we similarly show $\sum_{tr=1.5 \AA}^\infty P_n(l,tr)$ for a mutated AVV P5. The mutation, which has severe consequences for the 
promoter's ability to induce transcription (see Ref. \cite{Choi_2004}), consists of changing base-pairs $+1$ and $+2$ to $G-C$ pairs. From Fig. \ref{fig:3} 
we observe that 
this mutation indeed severely inhibits the formation of large bubbles around the $+1$ base-pair. Comparing Figs. \ref{fig:2} and \ref{fig:3}, 
we see that changing just two base pairs is a sufficient increase of the $G-C$ ``barrier'' to restrict the flow of thermal energy to be 
exclusively downstream in the sequence.
This is the mechanism by which the mutation can induce rather long-range effect, such as the  change in the probability around base pair $-30$, although this 
effect may be specific to periodic boundary conditions.

From these simulations we confirm that, for these heterogeneous sequences, the maxima of the PDFs 
correspond to biologically important sites in the sequence, and that even small changes of the sequence 
can lead to significant changes in the spatio-temporal probabilities.

In order to shed more light on the role of the life-time of the bubbles, we calculate 
a distribution function for the average bubble duration (ABD): 
\begin{eqnarray}
\ \textmd{ABD}\ (n,l,{tr})& = & \left\langle \frac{
\sum_{q_{n}^k=1}^{q_{n}^{k_{max}}(l,{tr})}\Delta t[q_{n}^k(l,{tr})]}{\sum_{q_{n}^k=1}^{q_{n}^{k_{max}}(l,{tr})}q_{n}^k(l,{tr})}
\right\rangle_M 
\end{eqnarray}
where the denominator
is the total number of bubbles, with strand-separation $tr$, in the $k^{th}$
simulation, spanning $l$ base pairs beginning at the $n$'th base-pair. It is important to emphasize that the 
information contained in the ABD can not be accessed through any thermodynamic considerations. 
In Fig. \ref{fig:4} we show the quantity $\sum_{tr=4.5 \AA}^ \infty ABD(n,l,tr)$ for the 
AAV P5 sequence as well as for the mutated AAV P5 sequence. 
The immediate observation is that the wild version of the AVV P5 sequence 
overall supports bubbles of significantly longer duration  than the mutated version. This is particularly 
true for bubbles of large strand separation. As documented by Fig. \ref{fig:3}, the mutated AAV P5 certainly 
supports a number of large bubbles but their duration is significantly shorter. Also, Fig. \ref{fig:4} shows 
that the region around base-pair $+1$ in the wild sequence supports large amplitude, long lived bubbles, a feature 
that is completely absent in the mutated sequence. 

\begin{figure}[h]
   \includegraphics[width=\columnwidth]{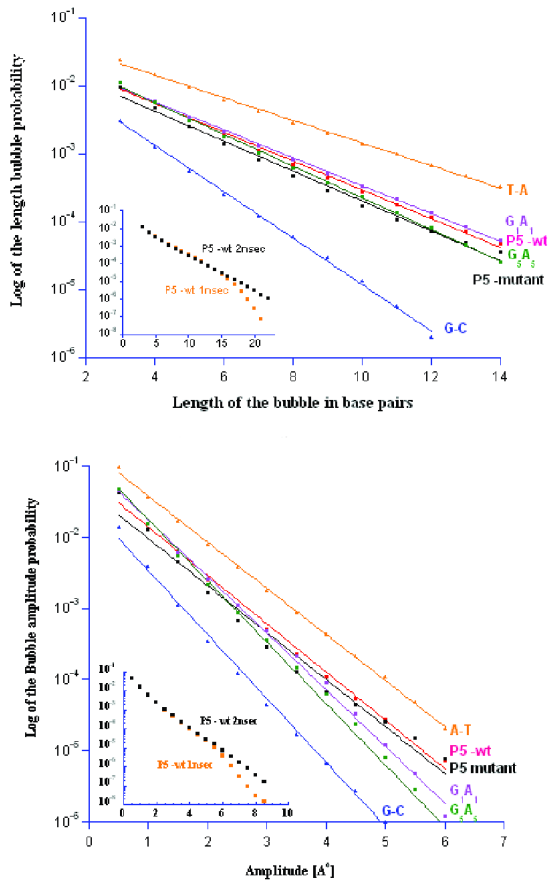}
   \caption{Bubble length probability (log scale), $\sum_{n=-46}^{23}\sum_{tr=1.5 \AA}^\infty P_n(l,tr)$ (upper), and bubble amplitude probability 
(log scale), $\sum_{n=-46}^{23} \sum_{l=3bp}^\infty P_n(l,tr)$ (lower). The insets compare these for AAV P5 
at 1 nsec and  2 nsec simulation times.} 
\label{fig:5}
   
\end{figure}
To compare the probability for bubble existence for all simulated sequences, we 
show in Fig. \ref{fig:5} the quantities $\sum_{n=-46}^{23}\sum_{tr=1.5 \AA}^\infty P_n(l,tr)$ (upper) and 
$\sum_{n=-46}^{23} \sum_{l=3bp}^\infty P_n(l,tr)$ (lower). 
In the two plots we show results for homogeneous $A-T$ and $G-C$ sequences, together with  AAV P5 and 
its mutated version. Also shown are results for the two periodic sequences $(G_1A_1)$ and $(G_5A_5)$. All 
probabilities decrease exponentially with size (amplitude and length) rendering large bubbles 
rare dynamical events. As is natural given the softer $A-T$ potential, the 
probability for any bubbles is always largest in a homogeneous $A-T$ sequence and lowest in a homogeneous $G-C$ sequence.
Comparing the results for homogeneous, periodic and heterogeneous sequences, it is clear that the bubble probability depends 
very little on sequence, but mainly on the AT and GC content \cite{commentgamma}. 
Finally, we observe that the periodic $(G_5A_5)$ sequence has 
less probability for longer bubbles in comparison with the sequence with smaller period $(G_1A_1)$ and with the heterogeneous AAV P5 sequence,
confirming that the GC ``barriers'' and 
their impedance role is restricting energy flow in the sequence. 
 
In the case of the bubble amplitude probability, there is a strong dependence on the actual sequence. The heterogeneous AAV P5 sequence 
sustains high amplitude bubbles significantly better than the periodic sequences $(G_1A_1)$ and $(G_5A_5)$ with the same 
AT content. Even the mutated AAV P5 sequences with slightly less AT content than the periodic sequences $(G_1A_1)$ and $(G_5A_5)$ is more probable to sustain
bubbles with amplitudes over $4 \AA$. Therefore the amplitude of the bubbles is 
is sensitive to the exact sequence. In the heterogeneous sequences the probability for 
bubble with high amplitudes is larger than in the periodic sequences with the same or even a little less AT content. This is consistent with 
recent demonstration \cite{weber} of melting temperatures being sensitive to intra-sequence correlation rather than being simply determined by AT and GC content.

The insets of Fig. \ref{fig:5} compare the results of the 1 nsec and 2 nsec simulations for the 
AAV P5 sequence. Since the results are equivalent up to amplitudes larger that $6 \AA$ and bubble lengths larger than $14$ base-pairs, we conclude that in this sequence the 
finite time effects in Langevin simulations exists only beyond these amplitudes and lengths.

In summary we have performed Langevin simulations of the PBD model of DNA and 
confirmed earlier results regarding the sequence dependence of bubble formation in agreement with results obtained on a purely thermodynamic 
basis. However, we find that the dynamics more sharply delineates the 
regions active for thermal strand separation because the 
life-times of bubbles are directly accounted for. We 
find that the probability for larger bubbles (lengths and amplitudes) is higher for heterogeneous
than for periodic sequences with the same A-T content. The important role of the length of the $G-C$ ``barriers'' for 
bubble existence was identified. We find that the bubbles with maximum duration begin their existence at biologically significant sites, and that 
these bubble initiation sites are different for bubbles with different amplitudes. Finally, we found  a striking sensitivity of the bubble life-time on  
sequence. Therefore we suggest that DNA's ability to sustain bubbles in some regions is a result of competition between length scales 
arising from the nonlinearity and the sequence heterogeneity, and that this competition sensitively controls the bubble life-times. Since
specific biological function are likely to be aided by long-lived openings of specific sizes, this information 
regarding size-lifetime relationships is directly relevant.

B.S.A. would like to thank Professor Peter Littlewood for the useful discussions.
Research at Los Alamos National Laboratory is performed
under the auspices of the US Department of Energy under contract W-7405-ENG-36.


\begin{thebibliography}{99}
\bibitem{sequence} A A-T (G-C) base pair is modeled by the parameters, $D_{AT}=0.05eV$, and $a_{AT}=4.2\AA^{-1}$
($D_{GC}=0.075eV$, and $a_{GC}=6.9\AA^{-1}$).
\bibitem{commentgamma}In general, the coefficients, $\gamma_{sequence}$ of the
exponential decrease are well approximated by a mean field consideration:
$ \gamma_{sequence}= \frac{n_{AT}\gamma_{AT}+n_{GC}\gamma_{GC}}{n_{AT}+n_{GC}}$, where $n_{AT}$ ($n_{GC}$) is the
number of AT (GC) base pairs. $\gamma_{A-T}$ ($\gamma_{G-C}$) is the coefficient for the pure $A-T$ ($G-C$) sequences.
\end{thebibliography}
\end{document}